\documentclass[12pt]{iopart}

\usepackage{epsfig,graphicx,dcolumn}
\usepackage[colorlinks,linkcolor=blue,anchorcolor=green,citecolor=blue]{hyperref}%%xlpeng
\usepackage{enumerate} %%xlpeng

\begin{document}

\title{Temporal prediction of epidemic patterns in community networks}

\author{Xiao-Long Peng$^{1,2}$, Michael Small$^{2}$, Xin-Jian Xu$^{1,3}$\footnote{Author to whom any correspondence should be addressed.} and Xinchu Fu$^{1,3}$}

\address{$^{1}$ Department of Mathematics, Shanghai University,
Shanghai 200444, People's Republic of China\\
$^{2}$ School of Mathematics and Statistics, The University of Western Australia, Crawley, WA 6009, Australia\\
$^{3}$ Institute of Systems Science,
Shanghai University, Shanghai 200444, People's Republic of China\\
}

\ead{xinjxu@shu.edu.cn}

\begin{abstract}
Most previous studies of epidemic dynamics on complex networks
suppose that the disease will eventually stabilize at either a
disease-free state or an endemic one. In reality, however, some
epidemics always exhibit sporadic and recurrent behaviour in one
region because of the invasion from an endemic population elsewhere.
In this paper we address this issue and study a
susceptible-infected-susceptible epidemiological model on a network
consisting of two communities, where the disease is endemic in one
community but alternates between outbreaks and extinctions in the
other. We provide a detailed characterization of the temporal
dynamics of epidemic patterns in the latter community. In
particular, we investigate the time duration of both outbreak and
extinction, and the time interval between two consecutive
inter-community infections, as well as their frequency
distributions. Based on the mean-field theory, we theoretically
analyze these three timescales and their dependence on the average
node degree of each community, the transmission parameters, and the
number of inter-community links, which are in good agreement with
simulations, except when the probability of overlaps between
successive outbreaks is too large. These findings aid us in better
understanding the bursty nature of disease spreading in a local
community, and thereby suggesting effective time-dependent control
strategies.
\end{abstract}

\pacs{87.19.X-, 89.75.Hc, 89.75.Fb}

\maketitle

%%%%%%%%%%%%%%%%%%%%%%%%%%%%%%%%%%%%%%%%%%%%%%%%%%%%%%%%%%%%%%%%%%%%%%%%%%%%%%%%%%%%%%%%%%
\section{Introduction}

Many social, communication, and biological systems of current
interest to the scientific community take the form of networks ---
sets of nodes (or vertices) joined together in pairs by links (or
edges) \cite{AB02,Newman10} --- with wide practical applications
ranging from searching on the Internet \cite{PV04} to epidemic
modelling \cite{BBV08}. One of the most interesting features of the
network is the presence of community structure (or modularity),
i.e., the division of network nodes into groups such that there is a
higher density of links within groups than between them
\cite{GN02,RCCL04,Newman06,LN08,POM09,Fort10}. For instance,
communities in a social network could be real social groupings, by
interest or background; communities on the web could be pages on
related topics.

Recent studies on network-based epidemic modelling have taken into
account the effect of the community structure
\cite{LH05,HPL06,SJ10,KL11}. It has been reported that the community
has a great impact on the magnitude of an outbreak peak and the
final prevalence \cite{LSS12}. Notably, some epidemic models
concerning communities have been extended to metapopulation
\cite{CPV07,LBP13} and interconnected (alternatively described as
coupled or layered) networks \cite{SSB12,MGC13} --- networks
consisting of interconnected and interdependent sub-networks or
communities --- and have revealed that in the case of two weakly
coupled networks, a new stable state may emerge, in which the
disease is endemic in one network but neither becomes endemic nor
dies out in the other \cite{DHS12,SD13}. This observation opposes
most previous studies that share an implicit assumption that the
disease will finally enter a steady state, either endemic or
disease-free \cite{AM91}.

Moreover, it has been frequently observed that some real diseases,
particularly the common zoonoses (diseases capable of cross-species
transmission), often trigger sporadic and recurrent human
infections. Zoonotic pathogens, in particular, are the major source
of emerging and re-emerging infections in humans \cite{Greger07}.
For instance, the highly pathogenic H5N1 influenza A virus remains a
zoonotic infection and is an endemic in avian populations, while it
rarely infects humans and is currently unable to sustain
human-to-human transmission \cite{Ken12}. Therefore, the continuing
reservoir of circulating influenza among the bird population and the
direct contacts between birds and humans (e.g., workers in poultry
farms) promote potential re-emergence of human infection, posing the
threat of an influenza pandemic \cite{WW03}. In such cases, the
disease neither persists nor becomes extinct forever in the human
population. In one region (or community) the disease neither
persists nor permanently vanishes, rather the disease experiences
sporadic and repeated cycles of outbreak and extinction. The
infection breaks out as the result of intermittent transmission of
pathogen from outside via the inter-community links and then dies
out because the infection rate is lower than the epidemic threshold
in the local community.

Although such a phenomenon has been observed in \cite{DHS12,SD13},
those references are primarily an investigation of  the conditions
necessary for the existence of such solutions. Neither the
associated timescales nor the frequencies of outbreak and extinction
have been satisfactorily investigated. Nevertheless, a detailed
knowledge of the temporal patterns of an epidemic in a local
community contributes to more sufficient preparedness in the face of
potentially pandemic disease \cite{WW03}. It is therefore of great
importance to properly investigate how frequently such outbreaks and
extinctions will happen, how long a single outbreak and extinction
will last, and how long it takes for the necessary inter-community
infection to occur. In this paper we address these problems by
studying a susceptible-infected-susceptible (SIS) model \cite{PV01}
on a random network with community structure. In particular, we
consider two interconnected communities with different average node
degrees, each of which is an Erd\"os-R\'enyi (ER) random graph
\cite{ER60}. We propose analytical predictions based on the
mean-field (MF) approach for these timescales and discuss their
dependence on the average node degree of each community, the
epidemiological parameters, and the number of inter-community links.
Furthermore, we test our analytical results against extensive
computational simulations and find good agreement --- particularly
when the probability of overlapping outbreaks is small. These
findings shed new light on the bursty nature of disease spreading
within a local community, which may help devise more efficient
time-dependent containment policies.

The rest of this paper is organized as follows. In section
\ref{sec2} we describe the algorithm for generating a random network
with community structure and introduce the epidemic model.
Analytical predictions regarding the probability and the expected
value of time spans over the system model are given in section
\ref{sec3}, while the numerical results are illustrated in section
\ref{sec4}. We conclude this work in section \ref{sec5}.

%%%%%%%%%%%%%%%%%%%%%%%%%%%%%%%%%%%%%%%%%%%%%%%%%%%%%%%%%%%%%%%%%%%%%%%%%%%%%%%%%%%%%%%%%%
\section{Model}\label{sec2}
\subsection{Network generation}
In the present work we consider the simple case of a random network
with two interconnected communities of different numbers of nodes
and different densities of links. However, it can be easily extended
to the general case that contains any number of communities of any
size.

The ER random graph \cite{ER60} is regularly used in the study of
complex networks, since networks with a complex topology and unknown
organizing principles often appear random \cite{AB02}. In this
paper, we generate a random network consisting of two interconnected
ER communities $\mathrm{A}$ and $\mathrm{B}$ of sizes $N_\mathrm{A}$
and $N_\mathrm{B}$ respectively. The following is the generation
process we adopt:

\begin{enumerate}[(i)]
  \item Assign each node to a single community, according to the communities' sizes.

  \item Generate ER community $\mathrm{A}$ ($\mathrm{B}$) with each pair of nodes in community $\mathrm{A}$ ($\mathrm{B}$) being connected with probability $p_\mathrm{A}$ ($p_\mathrm{B}$), following the standard construction procedures for ER random graphs \cite{ER60}.

  \item\label{previous} Join together by an inter-community link a randomly chosen node $i$ with intra-community degree larger than $1$, $k^{\mathrm A}_{i}>1$, in community $\mathrm{A}$ and a randomly selected node $j$ with intra-community degree larger than $1$, $k^{\mathrm B}_{j}>1$, in community $\mathrm{B}$. Here we specify the selections of nodes with respect to degree to ensure the definition of community in the strong sense \cite{RCCL04} --- that is, that each node of a community has more connections within the community than with the rest of the network. If a chosen node for inter-community connection in each community has already been connected by an inter-community link, do nothing and again randomly choose another node in the community.

  \item Repeat step (\ref{previous}) until there are $L$ inter-community links between communities $\mathrm{A}$ and $\mathrm{B}$.

\end{enumerate}

\begin{figure}
\includegraphics[width=\columnwidth]{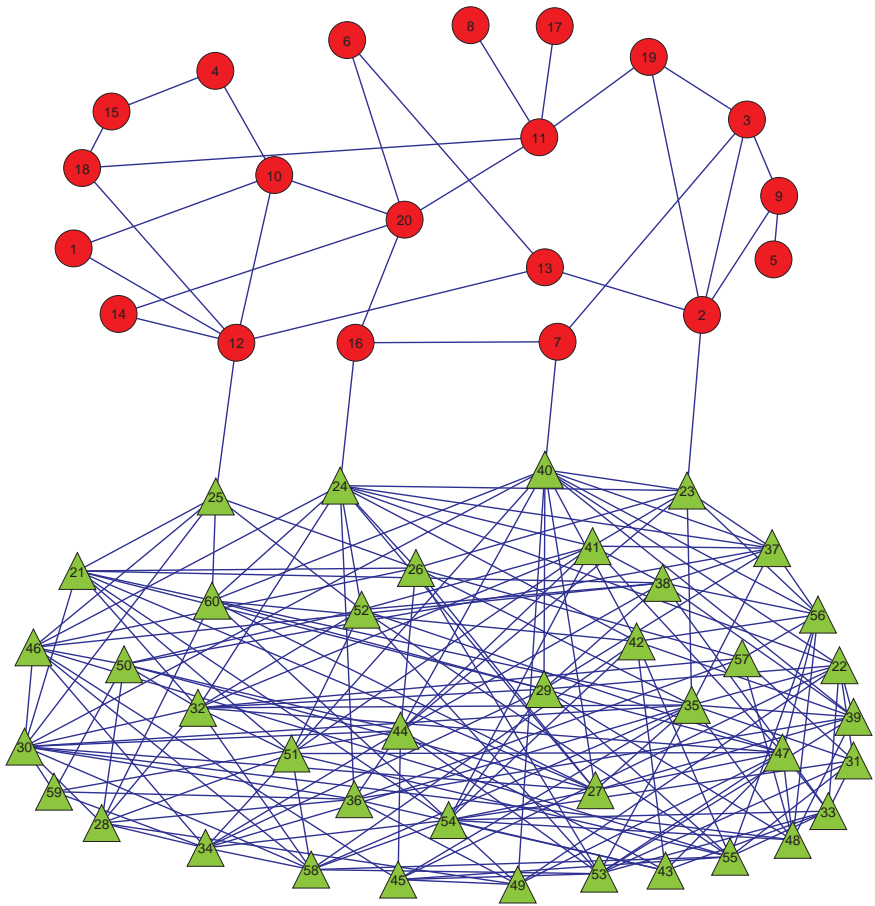}
\caption{(Colour online) A sample of random network with two
different communities, in which the sparser community
(${\mathrm{A}}$) consisting of 20 red nodes (denoted by circles) has
an average intra-community degree $\langle k_\mathrm{A}\rangle\simeq
3$ whereas for the denser community ($\mathrm{B}$) consisting of 40
green nodes (represented by triangles), $\langle
k_\mathrm{B}\rangle\simeq 10$. The number of inter-community links
between $\mathrm{A}$ and $\mathrm{B}$ is $L=4$.} \label{fig1}
\end{figure}

The above process generates a random network with community
structure, where each community has a Poisson distribution with
regard to the intra-community node degree. The average
intra-community node degrees of communities $\mathrm{A}$ and
$\mathrm{B}$ are then $\langle
k_\mathrm{A}\rangle=p_\mathrm{A}(N_{\mathrm{A}}-1)$ and $\langle
k_\mathrm{B}\rangle=p_{\mathrm{B}}(N_{\mathrm{B}}-1)$, respectively.
For analysis, we specifically  construct a relatively dense
community $\mathrm{B}$ and a sparse community $\mathrm{A}$ such that
$\langle k_\mathrm{B}\rangle$ is much larger than $\langle
k_\mathrm{A}\rangle$. According to step (\ref{previous}) both of the
endpoints of each inter-community link have an inter-community
degree $1$, $k_{\mathrm {AB}}=1$. A sample of random network with
community structure is illustrated in figure \ref{fig1}, where the
green (triangular nodes) community ($\mathrm {B}$) possesses a
higher density of connections than the red (circle nodes) one
($\mathrm {A}$).

\subsection{Epidemic model}

We consider the SIS epidemiological model taking place on the random
network generated above, where nodes mimic individuals or hosts and
links are the potentially infectious contacts among them. In this
model, each node is either susceptible (S) or infected (I).
Initially, there is a small fraction $\epsilon$ of infected nodes
only in the dense community ${\mathrm B}$ while all the remaining
nodes in the entire network are susceptible. Then at every time
step, each susceptible node is infected at a transmission rate
$\lambda$ upon a contact with an infected node. Meanwhile, each
infected node recovers and becomes susceptible again at a recovery
rate $\gamma$. For simplicity, we do not differentiate the
transmission parameters $\lambda$ and $\gamma$ for different
communities. That is, we assume that for each inter- or
intra-community link connecting an infected node with a susceptible
node in the entire network, the transmission rate $\lambda$ is
identical, and for each infected node in the entire network the
recovery rate $\gamma$ is the same. However, this can be easily
extended to the general case that assumes different transmission
rates for different communities, as in \cite{SSB12,SD13}.

\begin{figure}
\includegraphics[width=\columnwidth]{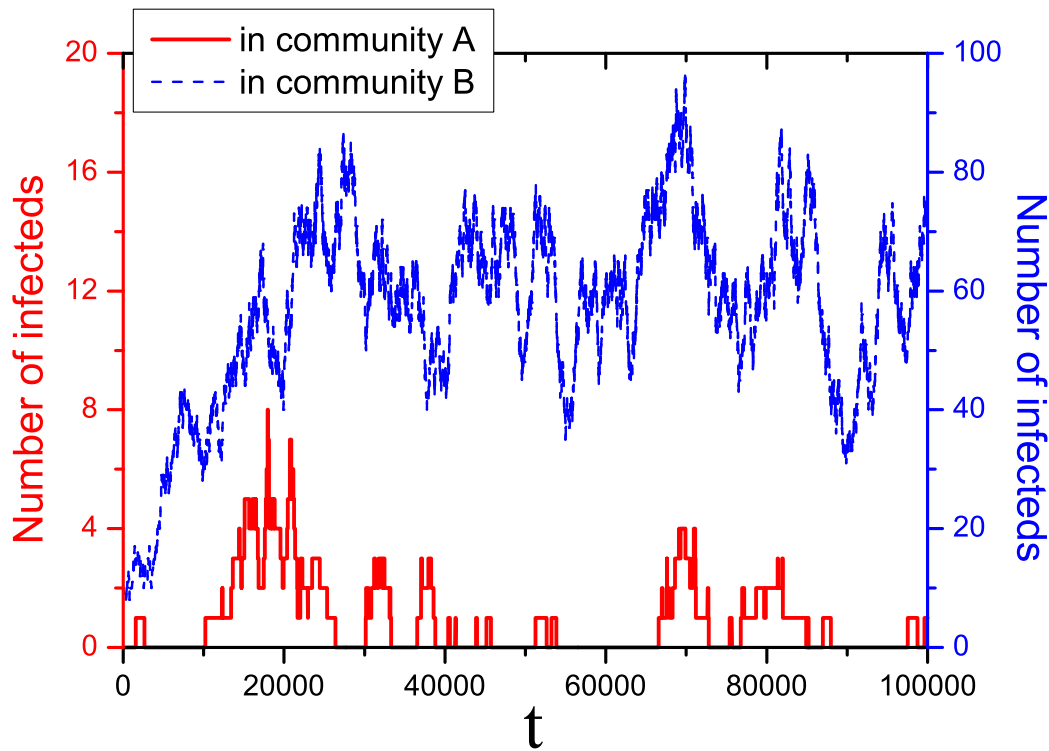}
\caption{(Colour online) A sample of time series of the number of
infected nodes in communities $\mathrm {A}$ (lower, red solid line)
and $\mathrm {B}$ (upper, blue dashed line), which are obtained by
simulating the model for at least $10^5$ time steps after initially
only infecting $10$ randomly selected nodes in community $\mathrm
{B}$. In the dense community $\mathrm {B}$, the disease persists for
the whole time window and the number of infected nodes fluctuates
around $60$ after an initial transient regime, whereas the sparse
community $\mathrm {A}$ experiences alternations between epidemic
outbreaks and extinctions. The parameters are: $N_{\mathrm
{A}}=100$, $N_{\mathrm {B}}=200$, $\langle k_{\mathrm
{A}}\rangle=4$, $\langle k_{\mathrm {B}}\rangle=12$, $L=10$,
$\lambda=0.001$, and $\gamma=0.008$, respectively.} \label{fig2}
\end{figure}

Of great importance in epidemic modelling is the basic reproductive
number $R_0$, which denotes the average number of infections
introduced by a single infected individual in a completely
susceptible population \cite{AM91}. This number characterizes the
threshold behaviour of a disease in the sense that it spreads across
a nonzero fraction of the population for $R_0>1$ while it dies out
for $R_0<1$. Note that it is easier for the disease to spread
through the denser community ${\mathrm B}$ than the sparser
community ${\mathrm A}$, since ${\mathrm B}$ has a larger basic
reproductive number than ${\mathrm A}$, i.e.,
$R_0^\mathrm{B}=\frac{\lambda}{\gamma}\langle k_\mathrm{B}\rangle$
$>\frac{\lambda}{\gamma}\langle k_\mathrm{A}\rangle=R_0^\mathrm{A}$
\cite{AM91}. The arguments for this expression of $R_0$ to hold in
the ER random networks, are twofold. First, the ER random graphs is
a typical example of homogeneous networks characterized by a Poisson
distribution $P(k)=\exp({-\langle k\rangle})\langle
k\rangle^{k}/{k!}$, where most node degrees are close to the average
degree, $k\simeq \langle k\rangle$. Therefore each node can be
assumed to have an identical number $\langle k\rangle$ of
neighbours. In accordance with the homogeneous assumption
\cite{BBV08}, one gets $R_0=\frac{\lambda}{\gamma}\langle k\rangle$.
Second, in a completely susceptible pool, the expected number of
susceptible neighbours that a newly infected node has is
$\sum_{k}\frac{k(k-1)P(k)}{\langle k\rangle}=\frac{\langle
k^2\rangle-\langle k\rangle}{\langle k\rangle}=\langle k\rangle$
\cite{DHS12}, thus according to the definition,
$R_0=\frac{\lambda}{\gamma}\langle k\rangle$. Here, the basic
reproductive number of each community is calculated based on the
respective average intra-community node degree, regardless of the
inter-community links, since this is the critical value for the
disease to spread within the single community. In addition, if
$R_0^\mathrm{B}>1$, $R_0^\mathrm{A}<1$, and $R_0^{\mathrm {Net}}>1$
(hereinafter, $R_0^{\mathrm {Net}}$ denotes the basic reproductive
number of the entire network), then the disease can spread and
persist within community ${\mathrm B}$, whereas in community
${\mathrm A}$ the disease might experience alternately outbreaks and
annihilations. As shown in figure \ref{fig2}, the number of infected
nodes in the sparse community fluctuates between zero and positive
values with time. This is because the disease is able to transmit
from community ${\mathrm B}$ (where the epidemic persists) to
community ${\mathrm A}$ through the inter-community links, which
play a pivotal role in introducing new infections to the otherwise
disease-free community. For analysis in the rest of this paper we
choose parameter values such that $R_0^\mathrm{B}>1$,
$R_0^\mathrm{A}<1$, and $R_0^{\mathrm {Net}}>1$.

Remarkably, an analogous phenomena have also been observed recently
in the study of epidemic spreading in interconnected or coupled
networks, with the emergence of a new stable state in which the
disease is endemic in one network but neither persists nor dies out
in the others \cite{DHS12,SD13}. Rather than focusing on the
conditions that permit this mixed phase \cite{DHS12,SD13}, in the
present work we probe further the timescales (temporal patterns) of
epidemic dynamics in the sparse community. In particular, we examine
the time durations of outbreaks and extinctions, and the time
interval between two successive inter-community infections, as well
as their frequency (probability) distributions. This issue has
seldom (if ever) been explored in the literature. However, the
direct answer can help us better understand how often infections and
extinctions will happen and how long a single outbreak and a single
extinction will last in a local community, and thus suggest more
effective time-dependent preventive measures.

%%%%%%%%%%%%%%%%%%%%%%%%%%%%%%%%%%%%%%%%%%%%%%%%%%%%%%%%%%%%%%%%%%%%
\section{Mean-field Analysis}\label{sec3}
Denoting by $x_{\mathrm{B}}(t)$ and $y_{\mathrm{B}}(t)$ the fraction
of susceptible and infected nodes in the dense community ${\mathrm
B}$ at time $t$, respectively, one has the normalization condition
$x_{\mathrm{B}}(t)+y_{\mathrm{B}}(t)=1$. Since the interconnections
between the weakly coupled networks only affect disease spreading in
the sparse network \cite{DHS12}, by disregarding the infections
along the inter-community links we can get the time evolution of the
fraction of infected nodes following the MF approach
\cite{PV01}:
\begin{equation}\label{eq1} \frac{\rmd}{\rmd
t}y_{\mathrm {B}}(t)= \lambda\langle k_{\mathrm {B}}\rangle
x_{\mathrm B}(t)y_{\mathrm {B}}(t)-\gamma y_{\mathrm {B}}(t),
\end{equation}
where the first term considers the infection of susceptible nodes
due to intra-community links, which is proportional to the
transmission rate $\lambda$, the average number $\langle k_{\mathrm
{B}}\rangle$ of intra-community neighbours per node in community
${\mathrm B}$, the density $x_{\mathrm {B}}(t)$ of susceptible
nodes, and the probability $y_{\mathrm {B}}(t)$ that a randomly
chosen intra-community neighbour is infected, while the second term
describes the recovery process of infected nodes, which is
proportional to the recovery rate $\gamma$ and the average density
$y_{\mathrm {B}}(t)$ of infected nodes. With the assumption
$R_0^\mathrm{B}>1$ and the stationary condition
$\frac{d}{dt}y_{\mathrm {B}}(t)=0$, it is straightforward to get the
nonzero solution in the steady state,
\begin{equation}\label{eq2}
y_{\mathrm {B}}=1-x_{\mathrm {B}}=1-\frac{\gamma}{\lambda\langle
k_{\mathrm {B}}\rangle},
\end{equation}
which is the epidemic prevalence in community $\mathrm {B}$. In
fact, the nontrivial analytical solution to (\ref{eq1}) can be
written as
\begin{equation}\label{eq3} y_{\mathrm
{B}}(t)=\frac{\lambda\langle k_{\mathrm
{B}}\rangle-\gamma}{\lambda\langle k_{\mathrm
{B}}\rangle-\exp{[-(\lambda\langle k_{\mathrm
{B}}\rangle-\gamma)(t-C_1)]}},
\end{equation}
where $C_1$ is a constant depending on the initial condition
$y_{\mathrm {B}}(0)=\epsilon$. As $t\rightarrow\infty$, (\ref{eq3})
again gives rise to the stable fixed point as in (\ref{eq2}). The
epidemic prevalence $y_{\mathrm {B}}$ can also be translated into
the probability that a randomly selected node in community $\mathrm
{B}$ becomes infected.

Let $q$ be the probability of disease spreading through the
inter-community links from community $\mathrm {B}$ into community
$\mathrm {A}$ in one time step. Consider that there are $m$ infected
nodes among the $L$ inter-community nodes in community $\mathrm {B}$
at a given time, with a probability ${L\choose m}{y^{m}_{\mathrm
{B}}}{(1-y_{\mathrm {B}})}^{L-m}$, and at least one inter-community
node in community $\mathrm {A}$ gets infected from the $m$
inter-community links with a probability $[1-(1-\lambda)^{m}]$.
Therefore, taking into account all possible numbers
($m=1,2,\ldots,L$) yields
\begin{equation}\label{eq4}
q=\sum_{m=1}^{L}{L\choose m}{y^{m}_{\mathrm {B}}}{(1-y_{\mathrm {B}})}^{L-m}[1-(1-\lambda)^{m}].
\end{equation}
If one defines $ T$ as the time interval between two consecutive
inter-community infections, then there is no inter-community
infection for a duration of $( T-1)$ time steps until the $ T$th
time step. As a result, the probability of the time span between two
successive inter-community infections being $ T$ time steps
is
\begin{equation}\label{eq5} P( T)=q(1-q)^{ T-1},
\end{equation}
and the average time interval between two consecutive inter-community infections is
\begin{equation}\label{eq6}
\langle  T\rangle=\sum_{ T=1}^{\infty}q(1-q)^{ T-1}  T.
\end{equation}
In fact, (\ref{eq4}) approximates to
\begin{equation}\label{eq14} q=\sum_{m=1}^{L}{L\choose
m}y_{\mathrm {B}}^{m}(1-y_{\mathrm {B}})^{L-m}m\lambda=y_{\mathrm
{B}}\lambda L
\end{equation}
if one neglects all higher order terms with regard to $\lambda$ in
the expression $[1-(1-\lambda)^m]$. Furthermore, taking $T$ as
continuous in (\ref{eq6}) yields
\begin{equation}\label{eq15}
\langle T\rangle =\int_{1}^{\infty} q(1-q)^{T-1}T\rmd
T=\frac{q[1-\ln(1-q)]}{[\ln(1-q)]^2},
\end{equation}
which can be simplified to $\langle T\rangle\approx 1/q$ for $q\ll
1$. This approximation treatment is feasible since the value of
$\lambda$ is set to be less than $0.01$ and the value of $\lambda L$
set to be up to $0.04$ for all simulations in this context.
Therefore the average time interval between two successive outbreaks
approximately scales as
\begin{equation}
\langle  T\rangle \approx \frac{1}{y_{\mathrm {B}}\lambda L} =
\frac{1}{L(\lambda-\gamma/\langle k_{\mathrm
{B}}\rangle)}.\label{eq16}
\end{equation}

\begin{figure}
%\begin{figure*}[htb]
%\centering
\includegraphics[width=\columnwidth]{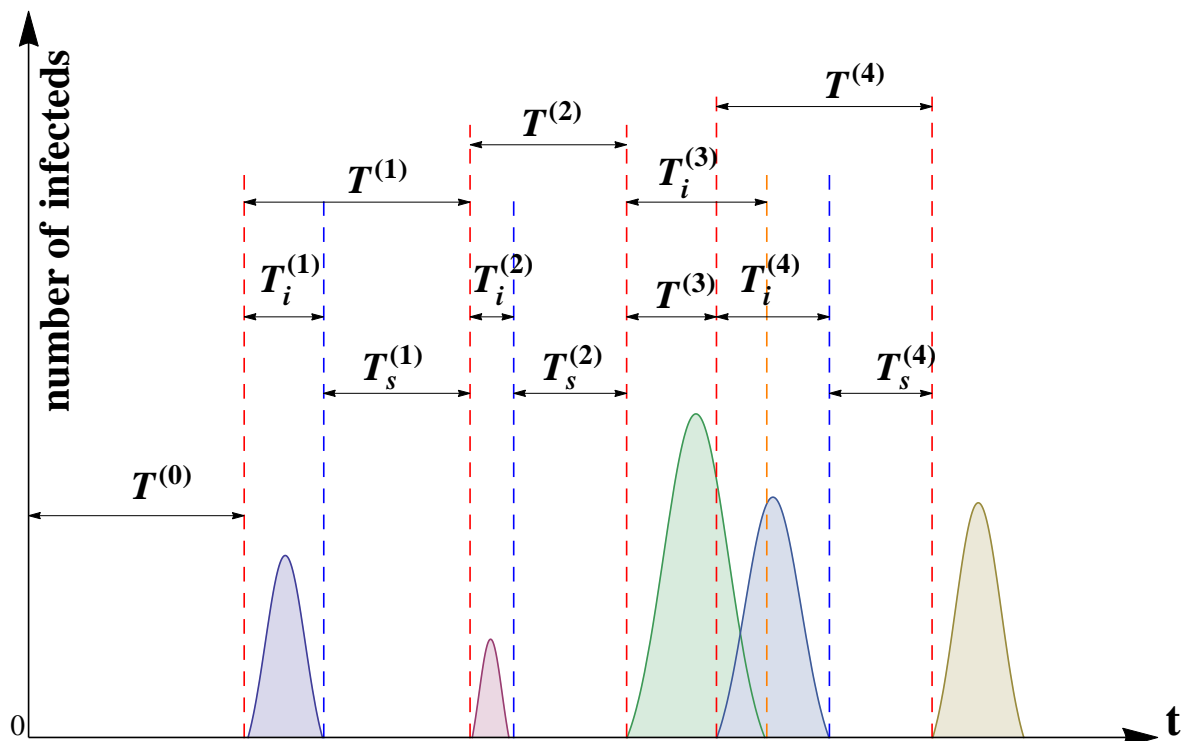}
\caption{(Colour online) Diagrammatic illustration for the
relationship between the three time durations: $T^{(j)}$,
$T_i^{(j)}$, and $T_s^{(j)}$, with the superscript $j$ denoting the
$j$th event of inter-community infection [excluding the first period
$T^{(0)}$], temporary epidemic outbreak and disease extinction in
community $\mathrm {A}$, respectively. For theoretical analysis, we
define an artificial (negative) time duration of disease extinction
[e.g., here, $T_s^{(3)}=T^{(3)}-T_i^{(3)}$] for the case of overlaps
between two successive epidemic outbreaks. Note that the curves on
top of the coloured areas do not represent the real number of
infected individuals, this is just for illustration.} \label{fig3}
%\end{figure*}
\end{figure}

As shown in figure \ref{fig2}, community $\mathrm {A}$ is free of
disease from the beginning until the epidemic from community
$\mathrm {B}$ transmits into this sparse group via the
inter-community links --- causing an epidemic outbreak. However the
infection can only last for a period of time, $ T_i$, which is
defined as the time duration (or span) of the outbreak, since the
basic reproductive number is too low ($R_0^{\mathrm {A}}<1$) for the
disease to persist in community ${\mathrm {A}}$. This means that for
a period of $ T_i$ time steps, the disease exists in community
${\mathrm {A}}$. After that community ${\mathrm {A}}$ is
disease-free for a period of time, $ T_s$, which is defined as the
time duration (or span) of the extinction (or health), with all the
nodes in the community being susceptible. This disease-free time
ends as the next inter-community infection succeeds. In detail, let
us assume that the disease initially comes into community $\mathrm
{A}$ after $ T^{(0)}$ time steps, and after that denote the series
of time intervals between consecutive inter-community infections by
$ T^{(1)}, T^{(2)},\ldots$, in sequence. Accordingly, we denote the
series of time durations of successive outbreaks by $ T^{(1)}_i,
T^{(2)}_i,\ldots$, and of successive extinctions by $ T^{(1)}_s,
T^{(2)}_s,\ldots$, see figure \ref{fig3} for an illustration.

It is complicated to derive theoretically the time spans $ T_i$ and
$ T_s$ due to the potential effects of inter-community infections
and the overlaps between successive epidemic outbreaks, which occur
when a new inter-community infection emerges before the old disease
dies out. However, there is a definite relationship between the
values of these three time spans. As seen in figure \ref{fig3}, if
two successive outbreaks do not overlap, then the time interval
between these two infections is equal to the sum of the time
duration of the former outbreak and the disease-free period before
the latter outbreak. Mathematically, that is
\begin{equation}\label{eq7}
T= T_i+ T_s\quad\quad \mathrm{if}\quad T> T_i.
\end{equation}
For example, from figure \ref{fig3} one can obtain $ T^{(1)}=
T_i^{(1)}+ T_s^{(1)}$, $ T^{(2)}= T_i^{(2)}+ T_s^{(2)}$, and $
T^{(4)}= T_i^{(4)}+ T_s^{(4)}$, but can not get a time duration of
health during the time interval of $ T^{(3)}$, since $ T^{(3)}<
T_i^{(3)}$. For analysis in theory, we specifically define a virtual
(negative) time span of health as $ T_s^{(j)}= T^{(j)}- T_i^{(j)}$
for the case of $ T^{(j)}< T_i^{(j)}$ associated to the
corresponding overlap. Such a treatment allows us to use the
relationship given in (\ref{eq7}) in the presence of overlaps
between consecutive outbreaks.

Once a node in community $\mathrm {A}$ gets infected by an
inter-community infection, it will trigger a temporal dynamical
process in its local community before a subsequent inter-community
infection occurs. Therefore, by omitting the effects of
inter-community transmissions, one can write the evolution equation
for $y_{\mathrm {A}}(t)$, the density of infected nodes in community
$\mathrm {A}$, similar to (\ref{eq1}), as
\begin{equation}\label{eq8}
\frac{d}{dt}y_{\mathrm {A}}(t)=(\lambda\langle k_{\mathrm
{A}}\rangle-\gamma) y_{\mathrm {A}}(t) - \lambda\langle k_{\mathrm
{A}}\rangle y^2_{\mathrm A}(t).
\end{equation}
Here, the second-order term arises from the normalization condition
where $x_{\mathrm A}(t)=1-y_{\mathrm A}(t)$ has been eliminated from
the equation. One can also work out the nontrivial solution to
(\ref{eq8}) which resembles (\ref{eq3}). Nonetheless, this can not
be used to evaluate the time duration of an outbreak because this
solution contains a constant closely related to the initial fraction
of infected individuals $y_{\mathrm {A}}(0)$, which is stochastic.
Since $\frac{d}{dt}y_{\mathrm {A}}(t)<0$ for $R_0^{\mathrm {A}}<1$,
the density of infected individuals decreases with time and is
difficult to reach a relatively high level. Therefore, we neglect
the second-order term with regard to $y_{\mathrm {A}}(t)$ in
(\ref{eq8}) and obtain an approximate solution
\begin{equation}\label{eq9}
y_{\mathrm {A}}(t)\approx y_{\mathrm
{A}}(0)\exp{[-(\gamma-\lambda\langle k_{\mathrm {A}}\rangle)t]}.
\end{equation}
We estimate the time period of the outbreak by applying the concept
of half-life ($t_{1/2}$) \cite{halflife}, generally defined as the
time required for a quantity to diminish to half its value as
measured at the beginning of the time period, which is typically
used to describe a quantity that follows an exponential decay.
Setting $y_{\mathrm {A}}(t)/y_{\mathrm {A}}(0)=1/2$, one gets the
half-life
\begin{equation}\label{eq10}
t_{1/2}=\frac{\ln2}{\gamma-\lambda
\langle k_{\mathrm {A}}\rangle}
\end{equation}
for the density of infected individuals. After  $n$ half-lives, the
density drops from $y_{\mathrm {A}}(0)$ to $y_{\mathrm
{A}}(0)/{2^n}$. In the present work we use the time, which is
required for the number of infected nodes $I^{\mathrm
{A}}_t=y_{\mathrm {A}}(t)N_{\mathrm {A}}$ to drop from the initial
value $I^{\mathrm {A}}_0=y_{\mathrm {A}}(0)N_{\mathrm {A}}$ to
$\frac{1}{2}$ (where the disease is close to extinction), to
estimate the time duration $ T_i$ of the outbreak [Strictly this
accumulation of half-lives is not exactly the time span $ T_i$ for
the entire outbreak; this approximation, however, matches the
qualitative behaviour of $ T_i$ and can provide a similar scaling
shape (if not the size), as shown in section \ref{sec4}.]. Thus,
letting $I^{\mathrm {A}}_0/{2^n}=\frac{1}{2}$ gives rise to
$n=1+\log_2{I^{\mathrm {A}}_0}$. Consequently, the time duration $
T_i$ is dependent on the initial number of infected nodes in
community $\mathrm {A}$ in such a way that $ T_i(I^{\mathrm
{A}}_0)=nt_{1/2}=(1+\log_2{I^{\mathrm {A}}_0})t_{1/2}$. Since there
are a number $L$ of inter-community links between the two
communities, all possible numbers of infected nodes at the beginning
of the outbreak period are $I^{\mathrm {A}}_0=1,2,\ldots,L$ as these
$I^{\mathrm {A}}_0$ inter-community links pass the disease
simultaneously with the probability
\begin{equation}\label{eq11}
P(I^{\mathrm {A}}_0)=\sum_{\ell=I^{\mathrm {A}}_0}^L {L\choose
\ell}y_{\mathrm {B}}^{\ell}(1-y_{\mathrm {B}})^{L-\ell}{\ell \choose
I^{\mathrm {A}}_0}\lambda^{I^{\mathrm
{A}}_0}(1-\lambda)^{\ell-I^{\mathrm {A}}_0}.
\end{equation}
By definition it is clear that $\sum_{I^{\mathrm
{A}}_0=1}^{L}P(I^{\mathrm {A}}_0)=q$, whereby we normalize
$P(I^{\mathrm {A}}_0)$ to obtain the average time duration of an
epidemic outbreak as follows:
\begin{equation}\label{eq12} \langle
T_i\rangle = \frac{1}{q}\sum_{I^{\mathrm {A}}_0=1}^{L} T_i
{(I^{\mathrm {A}}_0)} P(I^{\mathrm {A}}_0)
=\frac{\ln2}{q(\gamma-\lambda\langle k_{\mathrm
{A}}\rangle)}\sum_{I^{\mathrm {A}}_0=1}^{L}(1+\log_{2}{I^{\mathrm
{A}}_0})P(I^{\mathrm {A}}_0).
\end{equation}
For simplicity in theory, we estimate the average timescale of a single disease-free phase as
\begin{equation}\label{eq13}
\langle  T_s\rangle=\langle  T\rangle - \langle  T_i\rangle,
\end{equation}
using the expression in (\ref{eq7}), regardless of whether there are
overlaps between successive outbreaks.

Now, we analyze the frequency distributions $P(T)$ and $P(T_i)$ of
the timescales $T$ and $T_i$, respectively. In accordance with
(\ref{eq5}), one has
\begin{equation}\label{eq18}
P(T)\approx q\exp{(-qT)}
\end{equation}
for a negligibly small $q$, since
\begin{equation}\label{eq19}
(1-q)^T=\exp{(-qT)}-\frac{T}{2}q^2+o(q^2).
\end{equation}
The approximate solution (\ref{eq9}) implies that an epidemic
outbreak's probability of still existing at a time $t$ after its
first appearance is $\exp{[-(\gamma-\lambda\langle k_{\mathrm
{A}}\rangle)t]}$. Since the probability of disappearing during the
time interval $[t,t+\rmd t]$ is $(\gamma-\lambda\langle k_{\mathrm
{A}}\rangle)\exp{[-(\gamma-\lambda\langle k_{\mathrm
{A}}\rangle)t]}\rmd t$, we arrive at an exponential probability
distribution
\begin{equation}\label{eq20}
P(T_i)=(\gamma-\lambda\langle k_{\mathrm
{A}}\rangle)\exp{[-(\gamma-\lambda\langle k_{\mathrm
{A}}\rangle)T_i]}.
\end{equation}

\begin{figure*}[htb]
\centering
%\begin{figure}
\includegraphics[width=\textwidth]{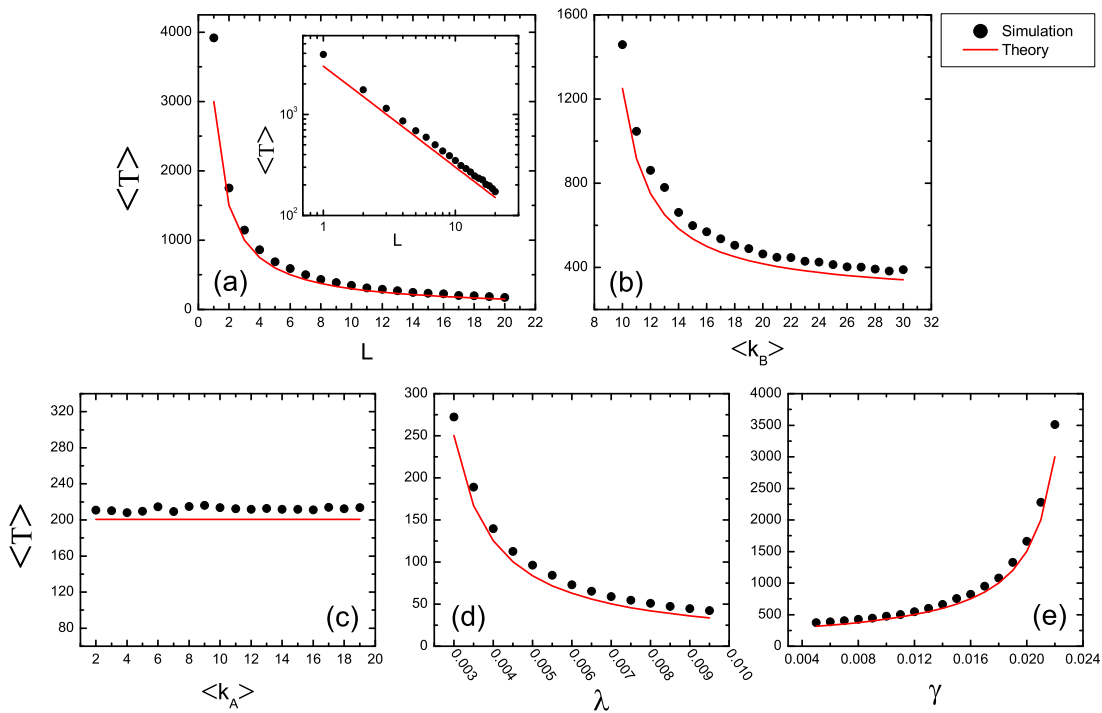}
\caption{(Colour online) The average time interval $\langle
T\rangle$ between two successive inter-community infections in
community $\mathrm {A}$ as a function of: the number $L$ of
inter-community links (with $\langle k_{\mathrm {A}}\rangle=4$,
$\langle k_{\mathrm{B}}\rangle=12$, $\lambda=0.001$, and
$\gamma=0.008$) (a); the average node degree $\langle k_{\mathrm
{B}}\rangle$ within community $\mathrm {B}$ (with $L=4$, $\langle
k_{\mathrm {A}}\rangle=4$, $\lambda=0.001$, and $\gamma=0.008$) (b);
the average node degree $\langle k_{\mathrm {A}}\rangle$ within
community $\mathrm {A}$ (with $L=20$, $\langle k_{\mathrm
{B}}\rangle=40$, $\lambda=0.01$, and $\gamma=0.0005$) (c); the
transmission rate $\lambda$ (with $L=4$, $\langle k_{\mathrm
{A}}\rangle=4$, $\langle k_{\mathrm {B}}\rangle=20$, and
$\gamma=0.04$) (d); and the recovery rate $\gamma$ (with $L=4$,
$\langle k_{\mathrm {A}}\rangle=4$, $\langle k_{\mathrm
{B}}\rangle=24$, and $\lambda=0.001$) (e), respectively. The log-log
plot of $\langle T\rangle$ in the inset of (a) exhibits a power law
$\langle T\rangle \propto L^{-1}$, which is consistent with
(\ref{eq16}). The black circles represent the simulation results and
the red lines denote the theoretical calculations by (\ref{eq6}).}
\label{fig4}
%\end{figure}
\end{figure*}

\begin{figure*}[htb]
\centering
\includegraphics[width=\textwidth]{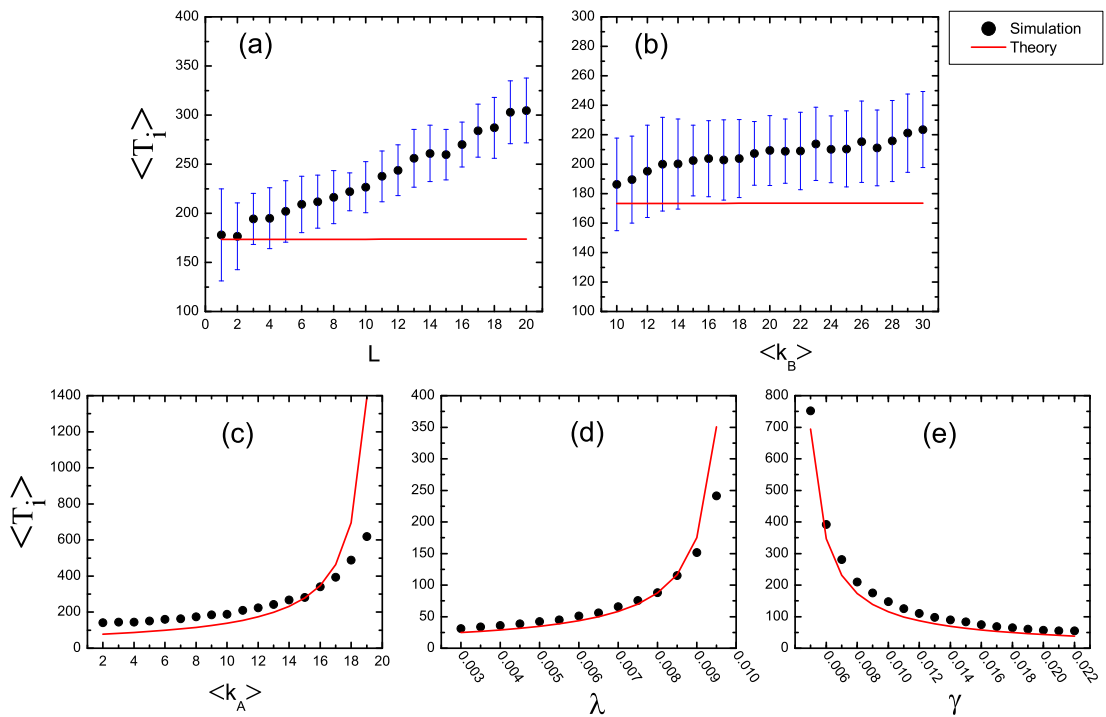}
\caption{(Colour online) The average time period $\langle
T_i\rangle$ of epidemic outbreak as a function of: the number $L$ of
inter-community links (with $\langle k_{\mathrm {A}}\rangle=4$,
$\langle k_{\mathrm{B}}\rangle=12$, $\lambda=0.001$, and
$\gamma=0.008$) (a); the average node degree $\langle k_{\mathrm
{B}}\rangle$ within community $\mathrm {B}$ (with $L=4$, $\langle
k_{\mathrm {A}}\rangle=4$, $\lambda=0.001$, and $\gamma=0.008$) (b);
the average node degree $\langle k_{\mathrm {A}}\rangle$ within
community $\mathrm {A}$ (with $L=20$, $\langle k_{\mathrm
{B}}\rangle=40$, $\lambda=0.01$, and $\gamma=0.0005$) (c); the
transmission rate $\lambda$ (with $L=4$, $\langle k_{\mathrm
{A}}\rangle=4$, $\langle k_{\mathrm {B}}\rangle=20$, and
$\gamma=0.04$) (d); and the recovery rate $\gamma$ (with $L=4$,
$\langle k_{\mathrm {A}}\rangle=4$, $\langle k_{\mathrm
{B}}\rangle=24$, and $\lambda=0.001$) (e), respectively. The black
circles represent the simulation results and the solid lines denote
the theoretical results obtained by (\ref{eq12}). The blue error
bars shown in the panels (a) and (b) are the standard deviation.}
\label{fig5}
\end{figure*}

\begin{figure*}[htb]
\centering
\includegraphics[width=\textwidth]{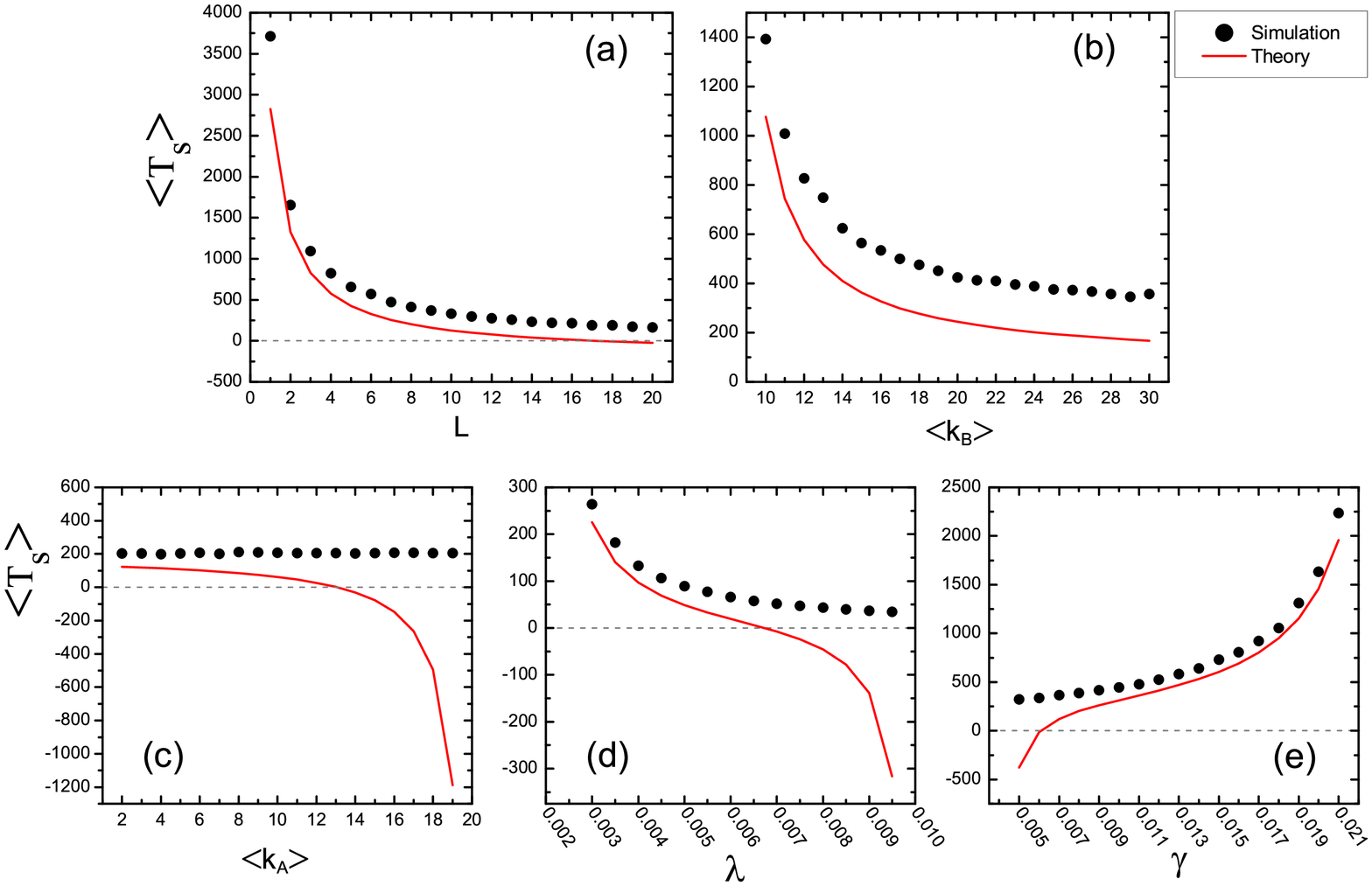}
\caption{(Colour online) The average time period $\langle
T_s\rangle$ of an extinction as a function of: the number $L$ of
inter-community links (with $\langle k_{\mathrm {A}}\rangle=4$,
$\langle k_{\mathrm{B}}\rangle=12$, $\lambda=0.001$, and
$\gamma=0.008$) (a); the average node degree $\langle k_{\mathrm
{B}}\rangle$ within community $\mathrm {B}$ (with $L=4$, $\langle
k_{\mathrm {A}}\rangle=4$, $\lambda=0.001$, and $\gamma=0.008$) (b);
the average node degree $\langle k_{\mathrm {A}}\rangle$ within
community $\mathrm {A}$ (with $L=20$, $\langle k_{\mathrm
{B}}\rangle=40$, $\lambda=0.01$, and $\gamma=0.0005$) (c); the
transmission rate $\lambda$ (with $L=4$, $\langle k_{\mathrm
{A}}\rangle=4$, $\langle k_{\mathrm {B}}\rangle=20$, and
$\gamma=0.04$) (d); and the recovery rate $\gamma$ (with $L=4$,
$\langle k_{\mathrm {A}}\rangle=4$, $\langle k_{\mathrm
{B}}\rangle=24$, and $\lambda=0.001$) (e), respectively. The black
circles represent the simulation results and the solid lines denote
the theoretical results obtained by (\ref{eq13}).} \label{fig6}
\end{figure*}

\begin{figure*}[htb]
\centering
\includegraphics[width=\textwidth]{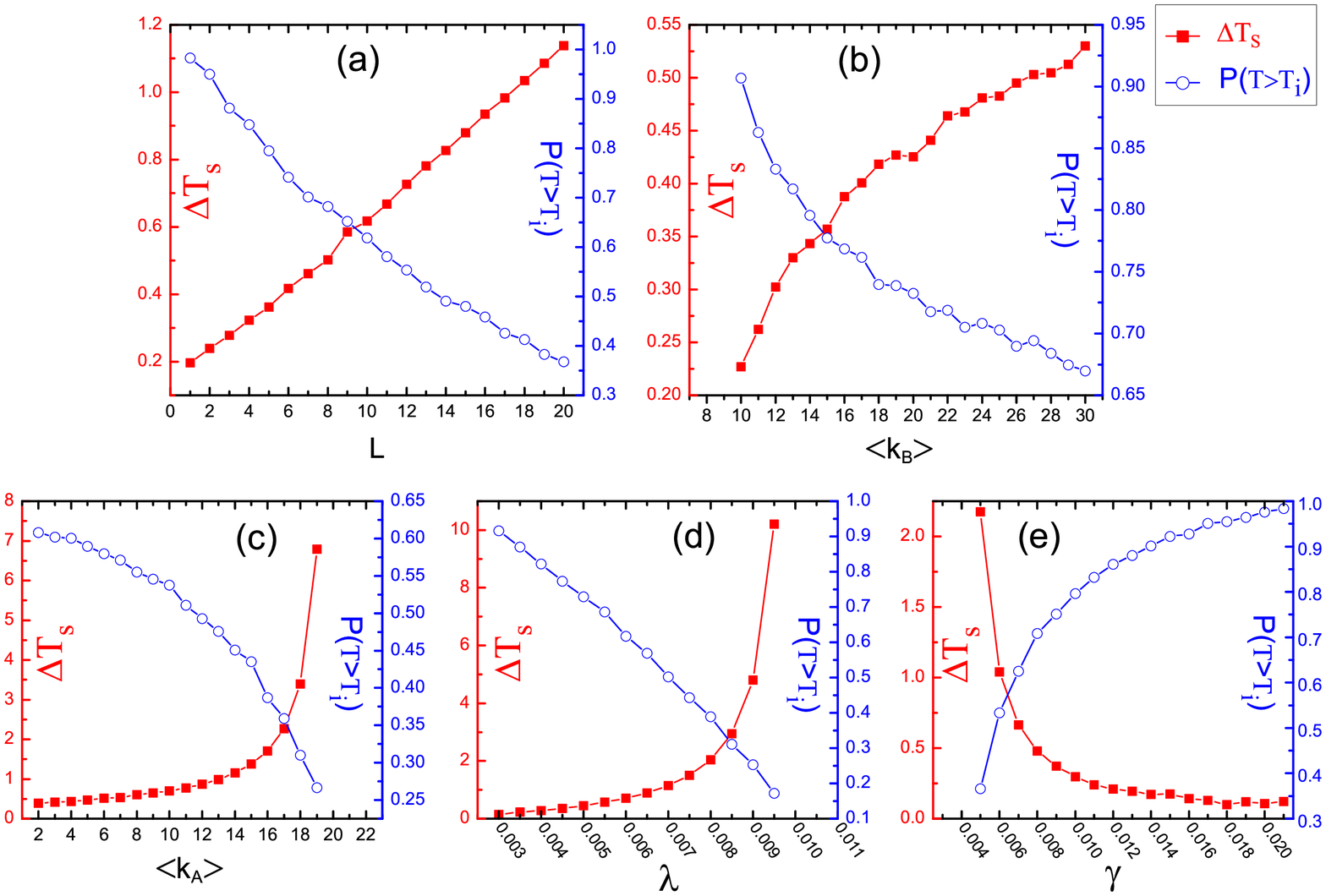}
\caption{(Colour online) Red squares represent the relative Errors
$\Delta T_s$ [defined by (\ref{eq17})] of the average disease-free
time period $\langle T_s\rangle$, of which the theoretical and
numerical results are correspondingly demonstrated in figure
\ref{fig6}. Blue circles denote the probability $P(T>T_i)$ of no
overlaps of epidemic outbreaks. All the results show that the fewer
overlaps between successive outbreaks, the more accurate theoretical
prediction.} \label{fig7}
\end{figure*}

\begin{figure}
\includegraphics[width=\columnwidth]{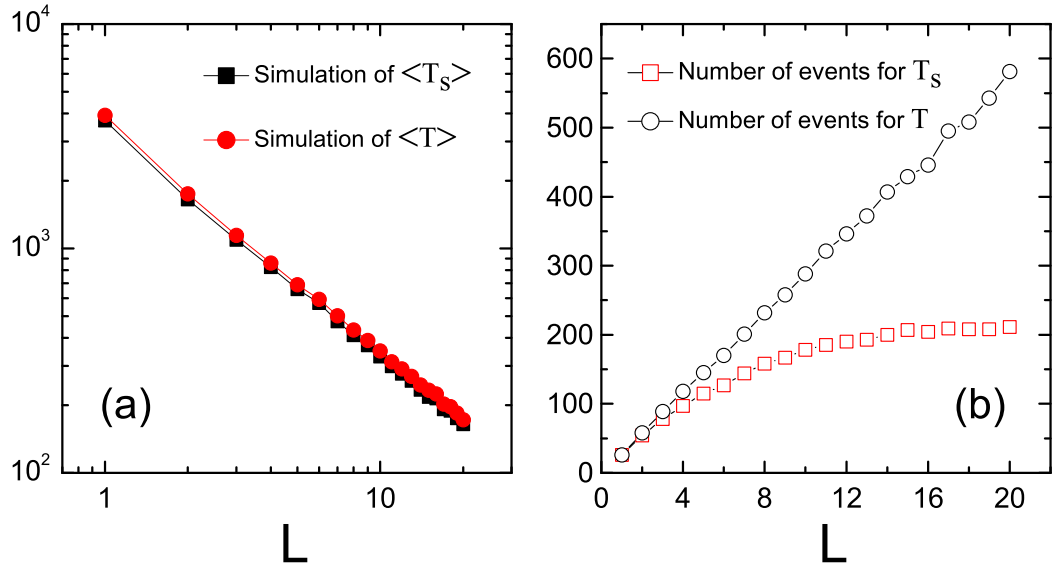}
\caption{(Colour online) Comparison of the simulation results (a) as
well as the numbers of events (b) between $\langle T\rangle$
(represented by circles) and $\langle T_s\rangle$ (represented by
squares) as the number of inter-community links $L$ varies. The
parameter values are the same as in figure \ref{fig4}(a).}
\label{fig8}\end{figure}

\begin{figure*}[htb]
\centering
\includegraphics[width=\textwidth]{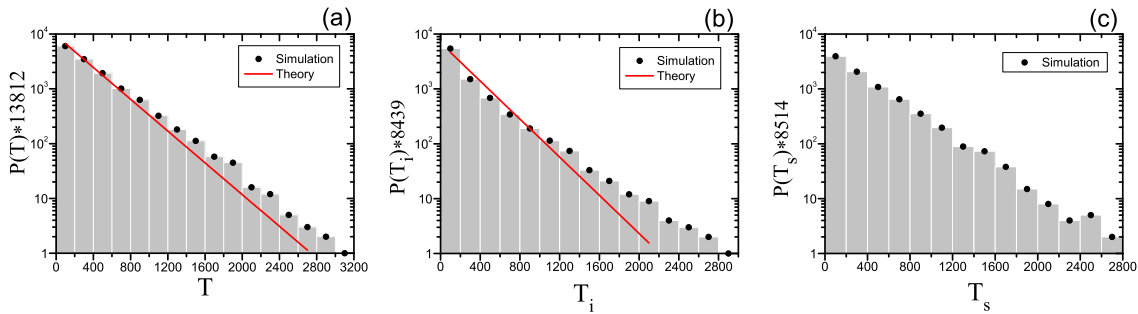}
\caption{(Colour online) Frequency distributions of:  $T$, the time
interval between successive inter-community infections (a); $T_i$,
the time duration of epidemic outbreak (b); and $T_s$, the time
period of disease-free (c). The simulation results shown in the
histogram are obtained after statistically counting the number of
events in each bin of width of $200$ time steps, with the solid
circles representing the counts of bin centers, while the red lines
in (a) and (b) represent the theoretical results obtained by
integrating Eqs.~(\ref{eq18}) and (\ref{eq20}) over each bin
respectively. Parameter values are $L=10$, $\langle k_{\mathrm
{A}}\rangle=4$, $\langle k_{\mathrm {B}}\rangle$, $\lambda=0.001$,
and $\gamma=0.008$. The total numbers of events for $T$, $T_i$, and
$T_s$ are $13812$, $8439$, and $8514$, respectively. } \label{fig9}
\end{figure*}

\begin{figure*}[htb]
\centering
\includegraphics[width=\textwidth]{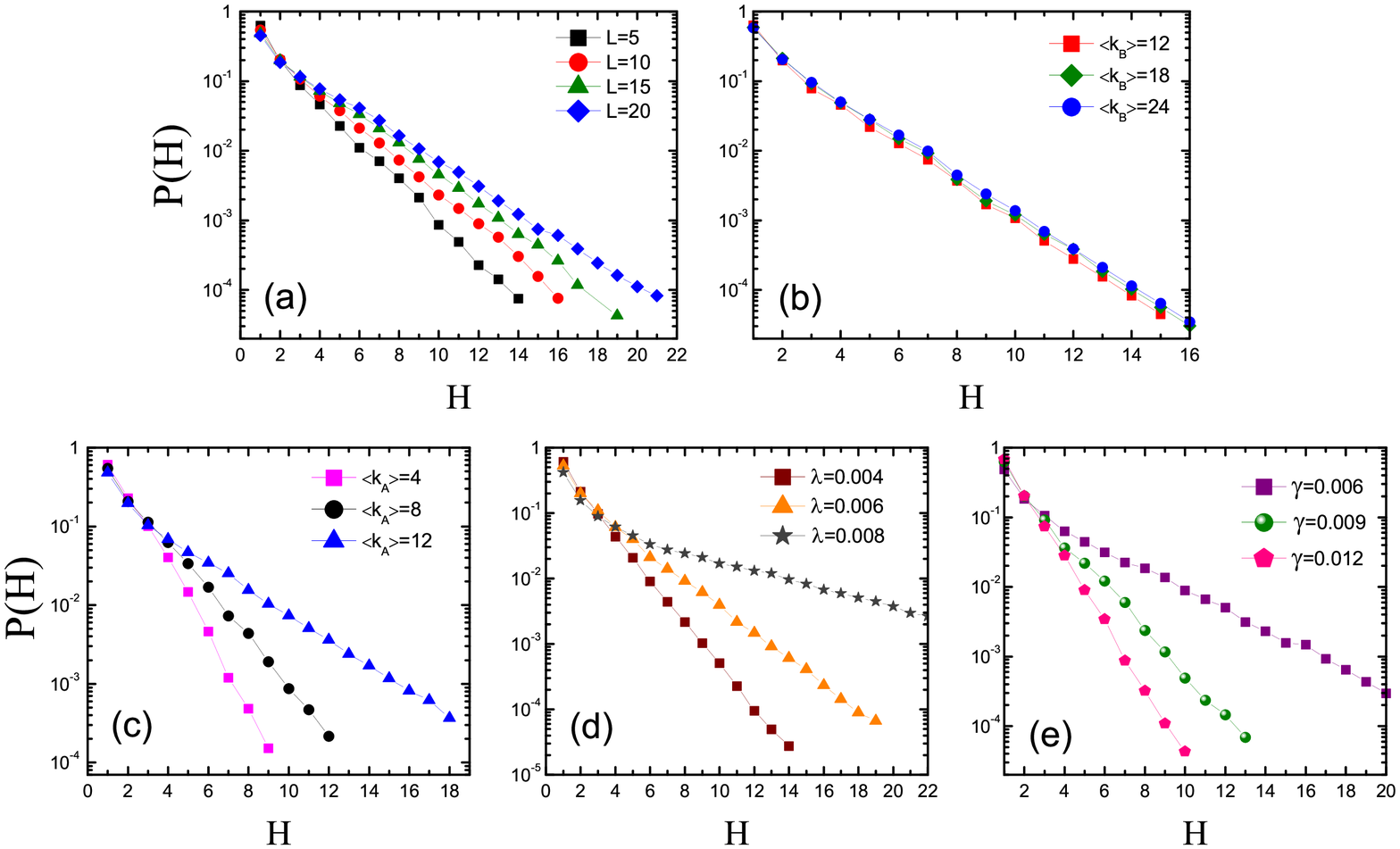}
\caption{(Colour online) Probability distribution $P(H)$ of the
height of epidemic outbreaks: for various values of $L$: 5, 10, 15,
20 (from bottom to top), with $\langle k_{\mathrm {A}}\rangle=4$,
$\langle k_{\mathrm{B}}\rangle=12$, $\lambda=0.001$, and
$\gamma=0.008$ (a); for various values of $\langle k_{\mathrm
{B}}\rangle$: 12, 18, 24 (from bottom to top), with $L=4$, $\langle
k_{\mathrm {A}}\rangle=4$, $\lambda=0.001$, and $\gamma=0.008$ (b);
for different values of $\langle k_{\mathrm {A}}\rangle$: 4, 8, 12
(from bottom to top), with $L=20$, $\langle k_{\mathrm
{B}}\rangle=40$, $\lambda=0.01$, and $\gamma=0.0005$ (c); for
various values of $\lambda$: 0.004, 0.006, 0.008 (from bottom to
top), with $L=4$, $\langle k_{\mathrm {A}}\rangle=4$, $\langle
k_{\mathrm {B}}\rangle=20$, and $\gamma=0.04$ (d); and for distinct
values of $\gamma$: 0.006, 0.009, 0.012 (from top to bottom), with
$L=4$, $\langle k_{\mathrm {A}}\rangle=4$, $\langle k_{\mathrm
{B}}\rangle=24$, and $\lambda=0.001$ (e), respectively.}
\label{fig10}
\end{figure*}

%%%%%%%%%%%%%%%%%%%%%%%%%%%%%%%%%%%%%%%%%%%%%%%%%%%%%%%%%%%%%%%%%%%%%%%%%%%%%%%%%%%%%%%%%%
\section{Results and discussions}\label{sec4}
\subsection{Average timescales}
We test theoretical predictions of the proviso section with
computational simulations of the epidemiological model over a random
network with community structure. We consider two ER communities
$\mathrm {A}$ and $\mathrm {B}$ with $N_{\mathrm {A}}=10^3$ and
$N_\mathrm {B}=2\times 10^3$ nodes, respectively. We start the
simulation with no infection in community $\mathrm {A}$ and a
fraction $\epsilon=0.5\%$ of infected nodes in community $\mathrm
{B}$, and then let the epidemic spreading go through $10^5$ time
steps for each realization. By recording the starting time of each
epidemic outbreak in community $\mathrm {A}$ arising from the
inter-community infections, we measure the time interval $T$ between
two consecutive outbreaks. We compute the time period, during which
at least one infected node exists in community $\mathrm {A}$, as the
time duration $T_i$ of an epidemic outbreak, and take the time
period, during which there is no disease in this community, as the
disease-free time duration $T_s$ of an extinction of disease. Note
that the time periods $T_i$ and $T_s$ calculated in simulations are
different from their theoretical estimates as long as there are
overlaps between epidemic outbreaks. See figure \ref{fig3} for
example, in case of an overlap between the third and the fourth
outbreaks, we numerically record one single epidemic outbreak with a
time period $[ T^{(3)}+ T_i^{(4)}]$ rather than two individual
outbreaks with time periods $T_i^{(3)}$ and $T_i^{(4)}$,
respectively. In addition, the numerical $T_s$ considers only the
positive time periods of extinction while disregarding the overlaps
of epidemic outbreaks.

To study the dependence of the average timescales on the network
structure and the transmission parameters, we calculate $\langle
T\rangle$, $\langle T_i\rangle$, and $\langle T_s\rangle$ by varying
each of the values of $L$, $\langle k_{\mathrm {A}}\rangle$,
$\langle k_\mathrm {B}\rangle$, $\lambda$, and $\gamma$ in such a
proper range that ensures $R_0^{\mathrm {A}}<1$, $R_0^{\mathrm
{B}}>1$, and $R_0^{\mathrm {Net}}>1$. All the simulation results
shown in the following figures are calculated averaging over at
least $20$ different initial network configurations, each performed
on $50$ realizations of the epidemiological model.

Figure \ref{fig4} shows the average time interval $\langle T\rangle$
between two successive epidemic outbreaks in community $\mathrm {A}$
as a function of the number $L$ of inter-community links [see figure
\ref{fig4}(a)], the average intra-community node degree $\langle
k_{\mathrm {B}}\rangle$ of community $\mathrm {B}$ [see figure
\ref{fig4}(b)], the average intra-community node degree $\langle
k_{\mathrm {A}}\rangle$ of community $\mathrm {A}$ [see figure
\ref{fig4}(c)], the transmission rate $\lambda$ [see figure
\ref{fig4}(d)], and the recovery rate $\gamma$ [see figure
\ref{fig4}(e)], respectively. On one hand, $\langle T\rangle$
decreases as $L$, $\langle k_{\mathrm {B}}\rangle$, and $\lambda$
increase. On the other hand, $\langle T\rangle$ increases with
$\gamma$, while keeping unchanged for varying values of $\langle
k_{\mathrm {A}}\rangle$. The more the inter-community links (or the
denser the community $\mathrm {B}$ or the larger the infection rate
or the smaller the recovery rate), the easier the inter-community
infections (i.e., the easier for the disease to spread from
community $\mathrm {B}$ into community $\mathrm {A}$), which
suggests a shorter time period between two successive epidemic
outbreaks. On the contrary, the property $\langle k_{\mathrm
{A}}\rangle$ inherent in the sparse community $\mathrm {A}$ has no
contribution to the inter-community infections which originate from
the dense community $\mathrm {B}$. Such behaivour of $\langle
T\rangle$ related to each of these parameters, including the
linearity shown in the log-log plot in the inset of figure
\ref{fig4}(a), confirms the MF analysis (\ref{eq16}) in section
\ref{sec3}. All the theoretical results by the MF approach are smaller than the simulations, since the MF approximation in (\ref{eq1}) has ignored the higher-order terms of $y_{\mathrm B}$, which leads to an overestimate for $y_{\mathrm B}$, and thus causes $\langle T\rangle$ to be underestimated according to (\ref{eq14}) and (\ref{eq16}).

Figure \ref{fig5} shows the mean time period $\langle T_i\rangle$ of
an epidemic outbreak in community $\mathrm {A}$ versus parameters
$L$, $\langle k_{\mathrm {B}}\rangle$, $\langle k_{\mathrm
{A}}\rangle$, $\lambda$, and $\gamma$, respectively. The theoretical
prediction of $\langle T_i\rangle$ remains almost unchanged with the
increase of $L$, whereas the simulation result shows a monotonous
increase. The reason for this is that the theoretical prediction is
based on the half-life calculation (\ref{eq10}) and only measures
part of the real period of the outbreak and also neglects the
effects of inter-community infections during the period of an
epidemic outbreak in the sparse community. In view of the fact that
the more the links between the two communities, the easier for the
inter-community transmission to happen. Thus, one would expect a
larger probability of overlaps between successive outbreaks [see
figure \ref{fig7}(a)], which causes numerical counting of $\langle
T_i\rangle$ to be more likely to exceed the theoretical prediction.
Therefore, it is hard to predict the average time period of an
epidemic outbreak if there are a large number of inter-community
links. A similar explanation can also be made for the difference
between the analysis and simulation in figure \ref{fig5}(b), where
the simulation results grow slightly with the increase of $\langle
k_{\mathrm {B}}\rangle$, whereas the analytical results are almost
unchanged. The higher density of connections inside community
$\mathrm {B}$ leads to a larger fraction of infected node in this
community in the steady state, which suggests a greater probability
of overlaps between epidemic outbreaks [see figure \ref{fig7}(b)].
However, the inter-community transmissions are heavily restricted
since the number of inter-community links is set to $L=4$, which can
explain why the simulation results of $\langle T_i\rangle$ increases
slowly compared to figure \ref{fig5}(a).

Moreover, we find from figure \ref{fig5}(c) and figure \ref{fig5}(d)
that both the analytical predictions and the simulation results of
$\langle T_i\rangle$ rise with the increase of $\langle k_{\mathrm
{A}}\rangle$ and $\lambda$, separately. For smaller values of
$\langle k_{\mathrm {A}}\rangle$ and $\lambda$, the predictions are
smaller than the simulations. This is due to the increase of
$\langle k_{\mathrm {A}}\rangle$ and $\lambda$ promoting the disease
spread and thus enhancing the possibility of overlapping outbreaks,
as seen in figure \ref{fig7}(c) and figure \ref{fig7}(d). However,
after the values of $\langle k_{\mathrm {A}}\rangle$ and $\lambda$
increase to a certain point, we see the reverse case, i.e., the
theoretical results are larger than the numerical results. It stems
from the simplification of (\ref{eq8}) by discarding the second
order term in $y_{\mathrm {A}}$. This approximation treatment
generally produces relatively tiny errors if the second order term
is far less than the first order term with respect to $y_{\mathrm
{A}}$. On the other hand, when the values of $\langle k_{\mathrm
{A}}\rangle$ and $\lambda$ become large enough so that
$(\gamma-\lambda \langle k_{\mathrm {A}}\rangle)$ approaches zero
and hence the first term closes to the second term in (\ref{eq8})
even for a very low level of $y_{\mathrm {A}}$. In this case,
neglecting the higher order term in $y_{\mathrm {A}}$ will
considerably underestimate the decaying speed of $y_{\mathrm {A}}$
and hence greatly overestimate the outbreak duration. In addition,
as shown in figure \ref{fig5}(e), both analytically and numerically
the average time period of $\langle T_i\rangle$ decreases with the
recovery rate $\gamma$. With a higher recovery rate, the infected
nodes will recover faster before they are able to spread the disease
to susceptible nodes, therefore the epidemic outbreak will last for
a shorter time. The theoretical prediction is smaller than the
simulation result because the half-life approximation only captures
part of the real time period of $\langle T_i\rangle$.

Figure \ref{fig6} displays the average time period $\langle
T_s\rangle$ of an extinction (i.e., a disease-free period) versus
$L$ [see figure \ref{fig6}(a)]; $\langle k_{\mathrm {B}}\rangle$
[see figure \ref{fig6}(b)]; $\langle k_{\mathrm {A}}\rangle$ [see
figure \ref{fig6}(c)]; $\lambda$ [see figure \ref{fig6}(d)]; and
$\gamma$ [see figure \ref{fig6}(e)], respectively. For all
parameters, the theoretical predictions given by (\ref{eq13}) are
smaller than the simulation results. This is to be expected since
(\ref{eq13}) covers both the positive part $\sum_{T>T_i}
(T-T_i)P(T-T_i)$ and the negative part $\sum_{T<T_i}
(T-T_i)P(T-T_i)$ whereas the simulation counting considers only the
positive part by ignoring those situations which include an overlap
between successive epidemic outbreaks. The negative part accounts
for a larger proportion if the probability of overlapping outbreaks
is larger, which will greatly reduce the accuracy of the theoretical
estimates of $\langle T_s\rangle$.

\subsection{Relative errors of $\langle T_s\rangle$}
Based on the above discussion, the deviation between the theoretical
and simulation results of $\langle T_s\rangle$ decreases with the
probability $P(T>T_i)$ that the time interval between two successive
outbreaks is larger than the time duration of the former outbreak
(with $1-P(T>T_i)$ being actually the probability of overlapping
outbreaks). To confirm this, we plot in figure \ref{fig7} the
relative error $\Delta T_s$ of the mean time duration of disease
extinction $\langle T_s\rangle$, which is defined as
\begin{equation}\label{eq17}
\Delta T_s=\frac{\big |{\langle T_s
\rangle}_{\mathrm{simulation}}-{\langle T_s
\rangle}_{\mathrm{theory}}\big |}{{\langle T_s
\rangle}_{\mathrm{simulation}}}\end{equation}
and can help examine
the accuracy of the theoretical prediction of $\langle T_s\rangle$.
A smaller $\Delta T_s$ denotes a more accurate prediction of
$\langle T_s\rangle$. As demonstrated in figure \ref{fig7}, the
accuracy of the theoretical estimates of $\langle T_s\rangle$
decreases with the increase of $L$, $\langle k_{\mathrm
{B}}\rangle$, $\langle k_{\mathrm {A}}\rangle$, and $\lambda$,
respectively. It is attributed to a larger number of inter-community
links (or a larger average node degree within each community or a
larger transmission rate) leading to a larger likelihood of
overlapping outbreaks. In particular, for large values of $\langle
k_{\mathrm {A}}\rangle$ [see figure \ref{fig7}(c)] and $\lambda$
[see figure \ref{fig7}(d)] the accuracy of $\langle T_s\rangle$
becomes worse. Since $\langle k_{\mathrm {A}}\rangle$ or $\lambda$
is large enough, the theory of $\langle T_i\rangle$ is much larger
than its simulation, causing $\langle T\rangle-\langle T_i\rangle$
to decline rapidly and become negative. Conversely, the theoretical
prediction of $\langle T_s\rangle$ is more accurate for a larger
value of $\gamma$, as shown in figure \ref{fig7}(e).

In addition, we observe an interesting fact that for all the
parameters $L$, $\langle k_{\mathrm {B}}\rangle$, $\langle
k_{\mathrm {A}}\rangle$, $\lambda$, and $\gamma$, the simulation
results of $\langle T_s\rangle$ are very close to those of $\langle
T\rangle$. As an example we demonstrate a comparison between the
simulation results of $\langle T_s\rangle$ and $\langle T\rangle$ as
$L$ varies in figure \ref{fig8}(a), and the numbers of events for
both the inter-community infections and disease extinctions are also
reported in figure \ref{fig8}(b). For a smaller value of $L$, the
numbers of both events are closer, meaning that the frequency of
outbreak overlaps is smaller, thus the difference between $\langle
T\rangle$ and $\langle T_s\rangle$ approaches to $\langle
T_i\rangle$, which is significantly smaller compared to $\langle
T\rangle$ at the small point of $L$. On the other hand, for a larger
$L$, the more frequent overlaps of outbreaks offset the difference
between $\langle T\rangle$ and $\langle T_s\rangle$ by counting the
small $T$ that satisfies $T<T_i$. So, the simulation results of $T$
and $\langle T_s\rangle$ are close for the whole range of $L$.
Similar results are also found in the simulations for the other
parameters.

\subsection{Distributions of the epidemic timescales}
 It
is clear from figure \ref{fig9}(a) that both the theoretical and
numerical results of $P(T)$ follow an exponential distribution. Note
that the simulation results have been counted for the frequency
distribution of $T$ and plotted in a histogram with an equal bin of
width $200$ time steps. The theoretical results depicted by the red
line are obtained after multiplying the integral of (\ref{eq18})
over each bin by the total event count ($13812$ in this example).
Apparently, the theoretical results decay faster, with a steeper
slope ($-q$), than the simulation. As mentioned before, the MF
approximation in (\ref{eq1}) overestimates the value of $y_{\mathrm
{B}}$ and hence, according to (\ref{eq14}), the theoretical value of
$q$ is larger than the numerical one. As shown in figure
\ref{fig9}(b), both the theory and the simulation exhibit an
exponential frequency distribution of the time duration $T_i$ of an
epidemic outbreak. Analogous to figure \ref{fig9}(a), the simulation
results in figure \ref{fig9}(b) are plotted as a histogram, where
the theoretical points present on the red line are given by
integrating (\ref{eq20}) over each bin of width $200$ time steps and
then multiplying the outcomes by the total event count --- $8439$.
For small values of $T_i$ [see the range $0<T_i<400$ in figure
\ref{fig9}(b)], the simulation results of $P(T_i)$ decrease faster
than the analytical prediction by (\ref{eq20}) because the
approximation used in (\ref{eq9}) neglects the second order term
with regard to $y_{\mathrm {A}}$ in (\ref{eq8}) and thus
underestimates the decaying speed compared with the simulation
results. On the other hand, for larger values of $T_i$, the
simulation results decay slower than the theory, because the
frequent overlaps between consecutive epidemic outbreaks help extend
the outbreak duration in the simulation computing. Since both $P(T)$
and $P(T_i)$ are exponential, the relationship given by (\ref{eq7})
allows us to expect an exponential frequency distribution for $T_s$,
which is confirmed by the simulation results shown in the histogram
of figure \ref{fig9}(c).

\subsection{Distributions of the peak heights}
Inspired by the intermittent occurrences of epidemic
outbreaks in the sparse community $\mathrm {A}$, we further
numerically investigate the probability distribution $P(H)$ of the
``height'' of an epidemic outbreak $H$, which is defined as the
maximal number of infected nodes during the outbreak. The simulation
results of $P(H)$ for various values of each of the parameters $L$,
$k_{\mathrm {B}}$, $k_{\mathrm {A}}$, $\lambda$, and $\gamma$ are
shown in figure \ref{fig10}, and all decay exponentially. This
results from that the frequency distribution $P(T_i)$ of time
duration of an outbreak follows an exponential decay, and, the
longer the outbreak duration $T_i$ the larger the probability that
the outbreak has a large height $H$. Figure \ref{fig10}(a) reveals
that the more the inter-community links, the more possible for the
epidemic outbreak to climb to a large height. According to the
simulation results in figure \ref{fig5}(a), a larger $L$ suggests a
longer time duration $\langle T_i\rangle$ of an outbreak since it is
more likely to cause overlaps between successive outbreaks, thus it
is easier to obtain a large $H$. From figure \ref{fig10}(b) we
observe that for a fixed value of $H$, the value of $P(H)$ increases
very slightly as $k_{\mathrm {B}}$ increases from $12$ to $18$ and
further to $24$. It arises from the increment of $k_{\mathrm {B}}$
resulting in a slight growth of $\langle T_i\rangle$ in simulations
since the relative small number ($L=4$) of inter-community links
restricts the inter-community infections, as explained for figure
\ref{fig5}(b). Based on (\ref{eq20}), we expect a higher frequency
$P(T_i)$ for larger values of $\langle k_{\mathrm {A}}\rangle$ and
$\lambda$ or for a smaller value of $\gamma$.  Thus, one can expect
an increase of $P(H)$ with the increase of $\langle k_{\mathrm
{A}}\rangle$ and $\lambda$ or with the decrease of $\gamma$, both of
which is supported by simulation results [see figure
\ref{fig10}(c-e)].

%%%%%%%%%%%%%%%%%%%%%%%%%%%%%%%%%%%%%%%%%%%%%%%%%%%%%%%%%%%%%%%%%%%%%%%%%%%%%%%%%%%%%%%%%%
\section{Conclusion}\label{sec5}

We have studied the SIS model in a random network composed of a
dense community $\mathrm {B}$ and a sparse community $\mathrm {A}$,
where the disease persists throughout the dense community while in
the sparse community it alternates between temporary epidemic
outbreaks and extinctions. The model is particularly relevant for
disease transmission from a reservoir population and intermittent
outbreak in a secondary group, as is the case with many zoonotic or
emerging infectious diseases. We have developed a theoretical
framework and performed extensive computational simulations to
understand the interesting features of the epidemic dynamics in the
sparse community. In particular, we have explored the expected
values for the time durations of outbreak and extinction, the time
interval between two successive outbreaks, and their frequency
distributions. The results demonstrated how these timescales rely on
the parameters including the number $L$ of inter-community links,
the average node degree $\langle k_\mathrm {B}\rangle$ within
community $\mathrm {B}$ and $\langle k_{\mathrm {A}}\rangle$ within
community $\mathrm {A}$, the transmission rate $\lambda$, and the
recovery rate $\gamma$. The theoretical results are in good
agreement with simulations except when there are too frequent
overlaps between successive epidemic outbreaks, since such overlaps
extend the time duration of outbreak and make the simulation
computing deviate from the theory. We have also found an exponential
decay for the frequency distributions of these timescales, as well
as for the frequency distribution of the maximal number of
infections during an epidemic outbreak. All these results may
provide helpful insights for understanding the temporal patterns of
an epidemic in a local community and lead to draw up effective
time-based preventive strategies. For example, the temporal pattern
of inter-community infections may suggests we implement a periodic
vaccination plan on the inter-community nodes in the sparse
community after each time period $\langle T\rangle$. More
effectively, according to the exponential distribution of the time
interval $T$ between successive inter-community infections, we may
also consider exponentially distributed vaccination forces
$\nu\propto exp(-qT)$ on the inter-community nodes in the sparse
community, with regard to the waiting time $T$ since the first
appearance of infection in the dense community.

This paper only considers the simplest network topology, i.e., the
ER random graph, which is an important example of homogeneous
networks. Therefore, it is of great interest to extend the present
work to heterogeneous networks, which needs a deeper study since the
heterogeneities in node degrees make it more complicated to make an
accurate prediction for the epidemic timescales. It is worth notice that the MF theory provides a good approximation for the prediction of the average timescale of $T$ and its distribution $P(T)$, which largely benefits from the fact that in the dense ER community, every node has a high degree and so do the nearest neighbours of any given node, where the MF theory generally yield good approximations to the underlying dynamical process \cite{GMWPM12}. However, as shown in the present work, the ignorance of higher-order correlations with respect to infectious density in the MF equation can lead to a small deviation from the simulations.

%%%%%%%%%%%%%%%%%%%%%%%%%%%%%%%%%%%%%%%%%%%%%%%%%%%%%%%%%%%%%%%%%%%
%\section*{Acknowledgments}
\ack
This work was jointly supported by NSFC Grant 11072136, SHMEC Grant 13YZ007, and STCSM Grant 13ZR1416800. M. S. was supported by an Australian Research Council Future Fellowship (FT 110100896). X.-L. P. acknowledges the China Scholarship Council for financial support and the University of Western Australia (UWA)--Shanghai University (SHU) Joint Training Program for support of this research.

%%%%%%%%%%%%%%%%%%%%%%%%%%%%%%%%%%%%%%%%%%%%%%%%%%%%%%%%%%%%%%%%%%%


\section*{References}

\begin{thebibliography}{99}

\bibitem{AB02}
Albert R and Barab\'{a}si A L 2002 {\it Rev. Mod. Phys.} {\bf 74} 47

\bibitem{Newman10}
Newman M E J 2010 {\it Networks: An Introduction} (New York: Oxford University Press)

\bibitem{PV04}
Pastor-Satorras R and Vespignani A 2004 {\it Evolution and Structure of the Internet} (New York: Cambridge University Press)

\bibitem{BBV08}
Barrat A, Barth\'elemy M and Vespignani A 2008 {\it Dynamical Processes on Complex Networks} (New York: Cambridge University Press)

\bibitem{GN02}
Girvan M and Newman M E J 2002 {\it Proc. Natl. Acad. Sci. USA} {\bf 99} 7821

\bibitem{RCCL04}
Radicchi F, Castellano C, Cecconi F, Loreto V and Parisi D 2004 {\it Proc. Natl. Acad. Sci. USA} {\bf 101} 2658

\bibitem{Newman06}
Newman M E J 2006 {\it Proc. Natl. Acad. Sci. USA} {\bf 103} 8577

\bibitem{LN08}
Leicht E A and Newman M E J 2008 {\it Phys. Rev. Lett.} {\bf 100} 118703

\bibitem{POM09}
Porter M A, Onnela J-P and Mucha P J 2009 {\it Not. Am. Math. Soc.}
{\bf 56} 1082

\bibitem{Fort10}
Fortunato S 2010 {\it Phys. Rep.} {\bf 486} 75

\bibitem{LH05}
Liu Z and Hu B 2005 {\it Europhys. Lett.} {\bf 72} 315

\bibitem{HPL06}
Huang L, Park K and Lai Y C 2006 {\it Phys. Rev. E} {\bf 73} 035103(R)

\bibitem{SJ10}
Salath\'e M and Jones J H 2010 {\it PLoS Comput. Biol.} {\bf 6} e1000736

\bibitem{KL11}
Kitchovitch S and Li\`o P 2011 {\it PLoS ONE} {\bf 6} e22220

\bibitem{LSS12}
Lentz H H K, Selhorst T and Sokolov I M 2012 {\it Phys. Rev. E} {\bf 85} 066111

\bibitem{CPV07}
Collizza V, Pastor-Satorras R and Vespignani A 2007 {\it Nature Phys.} {\bf 3} 276

\bibitem{LBP13}
Liu S Y, Baronchelli A and Perra N 2013 {\it Phys. Rev. E} {\bf 87} 032805

\bibitem{SSB12}
Saumell-Mendiola A, Serrano M \'A and Bogu\~n\'a M 2012 {\it Phys. Rev. E} {\bf 86} 026106

\bibitem{MGC13}
Mills H L, Ganesh A and Colijin C 2013 {\it J. Theor. Biol.} {\bf 320} 47

\bibitem{DHS12}
Dickison M, Havlin S and Stanley H E 2012 {\it Phys. Rev. E} {\bf 85} 066109

\bibitem{SD13}
Shai S and Dobson S 2013 {\it Phys. Rev. E} {\bf 87} 042812

\bibitem{AM91}
Anderson R M and May R M 1991 {\it Infectious Diseases of Humans: Dynamics and Control} (Oxford: Oxford University Press)

\bibitem{Greger07}
Greger M 2007 {\it Crit. Rev. Microbiol.} {\bf 33} 243

\bibitem{Ken12}
Berns K I {\em et al.} 2012 {\it Nature} {\bf 482} 153

\bibitem{WW03}
Webby R J and Webster R G 2003 {\it Science} {\bf 302}, 1519

\bibitem{PV01}
Pastor-Satorras R and Vespignani A 2001 {\it Phys. Rev. E} {\bf 63} 066117

\bibitem{ER60}
Erd\"os P and R\'enyi A 1960 {\it Publ. Math. Inst. Hung. Acad. Sci.} {\bf 5} 17

\bibitem{halflife}
Refer to http://en.wikipedia.org/wiki/Half-life for the information of half-life.

\bibitem{GMWPM12}
Gleeson J P, Melnik S, Ward J A, Porter M A and Mucha P J 2012 {\it
Phys. Rev. E} {\bf 85} 026106



\end{thebibliography}
\end{document}